# Machine Learning Applications in Spine Biomechanics

Farshid Ghezelbash[1*] | Amir Hossein Eskandari[1,2] | Xavier Robert-Lachaine[2] | Frank Cao[3] |
Mehran Pesteie[4] | Zhuohua Qiao[5] | Aboulfazl Shirazi-Adl[1] | Christian Larivière[2]

## Abstract

Spine biomechanics is at a transformation with the advent and integration of machine learning and computer vision technologies. These novel techniques facilitate the estimation of 3D body shapes, anthropometrics, and kinematics from as simple as a single-camera image, making them more accessible and practical for a diverse range of applications. This study introduces a framework that merges these methodologies with traditional musculoskeletal modeling, enabling comprehensive analysis of spinal biomechanics during complex activities from a single camera. Additionally, we aim to evaluate their performance and limitations in spine biomechanics applications. The real-world applications explored in this study include assessment in workplace lifting, evaluation of whiplash injuries in car accidents, and biomechanical analysis in professional sports. Our results demonstrate potential and limitations of various algorithms in estimating body shape, kinematics, and conducting in-field biomechanical analyses. In industrial settings, the potential to utilize these new technologies for biomechanical risk assessments offers a pathway for preventive measures against back injuries. In sports activities, the proposed framework provides new opportunities for performance optimization, injury prevention, and rehabilitation. The application in forensic domain further underscores the wide-reaching implications of this technology. While certain limitations were identified, particularly in accuracy of predictions, complex interactions, and external load estimation, this study demonstrates their potential for advancement in spine biomechanics, heralding an optimistic future in both research and practical applications.

## 1 Introduction

Over the years, the requirement for complex systems with multiple sensors and video cameras to accurately measure a wide range of biomechanical parameters has limited their applications in workplace, sport, and recreational activities (Cronin, 2021; Kidziński et al., 2020). Specifically, carrying out precise, subject-specific analysis of the human spine has relied heavily on kinematics measurements and segmental body mass estimations that are vital as inputs for musculoskeletal models. However, in real-world applications such as sports, ergonomics, and forensic studies, obtaining these inputs is often a time-consuming and difficult task. For instance, accurately estimating biomechanical parameters in a workplace with the goal to assess and prevent spinal injuries can be challenging, as it demands wearing devices and sensors for long periods that may not only be exhausting but could interfere with the natural posture and movements; furthermore, in the case of inertial sensors, magnetic field could adversely affect their reliability (Robert-Lachaine et al., 2020). Likewise, in scenarios such as accident reconstructions where biomechanical analysis could be used as evidence in a court, the only available data may be a single piece of footage.

[1] Division of Applied Mechanics, Department of Mechanical Engineering, Polytechnique Montréal, Canada
[2] Institut de Recherche Robert Sauvé en Santé et en Sécurité du Travail, Montréal, Canada
[3] Department of Mechanical Engineering and Material Science, Duke University, USA
[4] Department of Electrical and Computer Engineering, University of British Columbia, Canada
[5] Department of Mechanical Engineering, McGill University, Canada
[*] ghezelbash.far@gmail.com

Recent advancements in machine learning and computer vision technologies have sparked a revolution in this field, making it possible to estimate multiple biomechanical parameters from a single camera image/video. As one of the most pivotal biomechanical parameters, kinematics was traditionally collected via complex multi-camera systems or gyroscopic sensors. Recent improvements in deep neural networks have led to the creation of pose estimation algorithms that can predict the 3D human pose using a single camera input (Bogo et al., 2016; Cao et al., 2017). However, while most efforts in biomechanics have focused on gait analysis (Kidziński et al., 2020; Stenum et al., 2021; Yamamoto et al., 2022) or specific tasks (Cronin, 2021; Haberkamp et al., 2022), there is a paucity in the application of these packages in spine biomechanics across a broad range of applications. Equally important are the developments in the area of segmental mass estimations, which could have substantial impacts on spine biomechanical analyses (Liu et al., 2023) – especially in obese individuals (Ghezelbash et al., 2017). Previously, these estimations depended on tools such as 3D scanning (Pryce and Kriellaars, 2014) or dual x-ray absorptiometry (DEXA) (Bachrach, 2000), but machine learning innovations have now provided the ability to estimate 3D body shapes from a single image (He et al., 2021; Saito et al., 2019), thus obviating the need for specialized equipment. Notwithstanding these significant advances, there remains a gap in the comprehensive development and exploration of the practical relevance and applicability of pose estimation algorithms and 3D scan estimations in biomechanics (Cronin, 2021). This opens up a new frontier for future research and developments, posing an important opportunity to advance the current understanding and applications of these tools.

In this study, we aim to integrate recent advancements in machine learning and computer vision technologies with a subject-specific musculoskeletal model of the spine to propose a novel biomechanical evaluation framework. In addition, we aim to assess the efficacy and constraints of these tools in spinal biomechanics, hypothesizing that integrating these advanced technologies with traditional biomechanical methods can enhance the analysis of spinal biomechanics from simple camera images/videos. This approach synergistically combines traditional biomechanical methodologies with innovative technologies to enable detailed analyses of spinal biomechanics from a single camera image/video. Recognizing the constraints of real-world applications in sports, ergonomics, and forensic studies, we address the challenge of estimating biomechanical parameters without the need for multiple sensors and multi-camera systems. 3D body shape estimations, we evaluate the performance of PIFuHd in reconstructing body shapes, waist circumference and predicting segmental masses; also, for pose estimation algorithms, we explore the performance and limitations of various pose estimation algorithms like Blazepose (Bazarevsky et al., 2020), ICON (Xiu et al., 2022), and MeTRAbs (Sárándi et al., 2023), exploring their applicability in diverse dynamic tasks such as repetitive lifting, asymmetric lifting, whiplash neck injury, deadlift, and snatch weightlifting. This study elucidates both novel opportunities and existing challenges in these approaches while demonstrating their potential for future in-field biomechanical analyses.

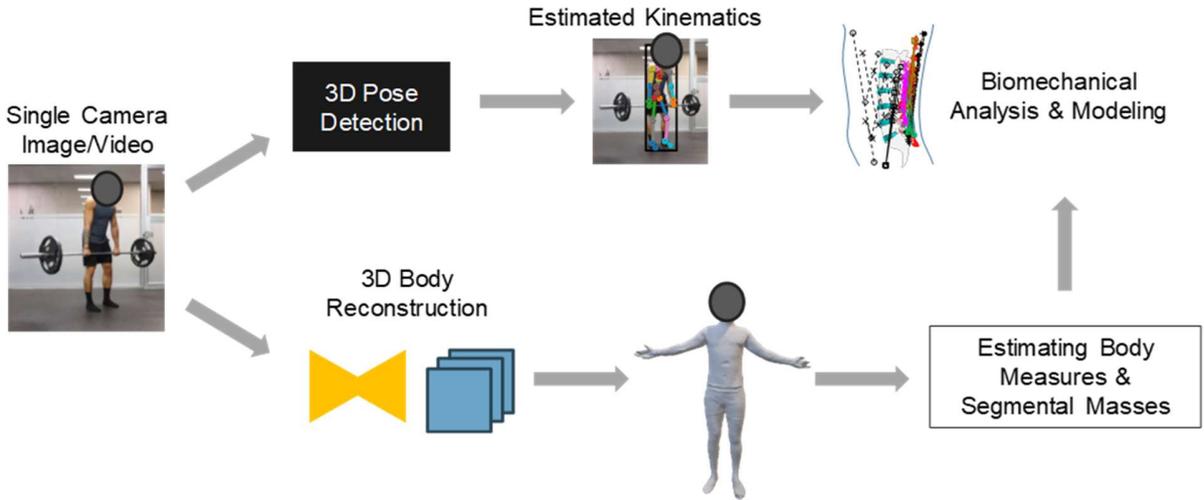

Figure 1: The outline of the proposed framework for integrating kinematics estimation, 3D body shape analysis, musculoskeletal modeling as well as biomechanical analysis of the spine from a single image.

## 2  Methods

To carry out a comprehensive biomechanical analysis with the focus on spine, we propose a framework that integrates three different approaches: kinematics estimation, 3D body shape analysis, as well as musculoskeletal analysis and modeling (Figure 1). Using a single image input, this framework utilizes deep neural network packages to estimate associated kinematics and body shape, which will drive a musculoskeletal model. This novel protocol, by combining these aspects into one cohesive system, facilitates a comprehensive biomechanical analysis of the spine. In the following section, we explore its accuracy and robustness via several applications.

## 2.1 Body Shape and Anthropometric Estimations

We collected images of four young lean males (average age = 31 years) from T-poses in the frontal plane. These images were then processed to estimate associated 3D body scans using PIFuHD (Saito et al., 2020). Initially, we compared the measured waist circumferences with those computed from the 3D scans. Furthermore, by assuming a uniform body density of 1.071 g/cm³ (Pryce and Kriellaars, 2014), we estimated the weight of the trunk that was subsequently compared with our results obtained from regression equations developed based on DEXA scans of 4,000 individuals (National Center for Health Statistics, 1999). These regression equations utilize weight, height and waist circumference to predict the trunk mass; for details on the developed regression equations, the following GitHub repository: https://github.com/amhoesk/body-fat-estimation.

## 2.2 Pose Estimation

For pose estimation, among packages available, we selected Blazepose (Bazarevsky et al., 2020), ICON (Xiu et al., 2022), and MeTRAbs (Sárándi et al., 2023) for this study. Initially, we evaluated the performance of these packages using the Fit3D database (Fieraru et al., 2021) in predicting the trunk flexion (defined as the angle between the vertical axis and a line connecting the mid-shoulder to mid-hip) during various exercises (i.e., barbell dead row, clean and press, barbell shrug, and deadlift) in nine subjects. Furthermore, we assessed the accuracy of these packages in predicting the trunk flexion angle in various applications. It should be noted that MeTRAbs (Sárándi et al., 2023), with the best performance among the selected packages and possessing the largest training dataset, was often our preferred choice for use in the subsequent analyses.

### 2.2.1 Application I: Repetitive Lifting and Lyapunov Exponent

We evaluated the performance of ICON (Xiu et al., 2022) and MeTRAbs (Sárándi et al., 2023) to estimate the flexion angle and corresponding Lyapunov exponent (as a stability metric) in a cyclic task. A male subject was instructed to repeatedly lift and lower a 4 kg load 35 times at the rate of 10 cycles per minute. The whole-body kinematics of the subject were measured using Xsens at 60 Hz, using 17 MTw Awinda inertial sensors (Xsens Motion Technologies, Enschede, The Netherlands). Additionally, a video of the subject was captured with a mobile phone camera and used to estimate the kinematics with ICON and MeTRAbs. Maximum Lyapunov exponents of the trunk flexion angle were determined using an existing algorithm (Graham and Brown, 2012; Rosenstein et al., 1993). The Lyapunov exponents were estimated as the mean rate of separation of the nearest neighbors, using a least-squares fit to the average. For further details on the Lyapunov exponent calculation and experiments, see (Eskandari et al., 2023b).

### 2.2.2  Application II: Asymmetric Lifting

Due to the prevalence of asymmetric lifting in occupational activities at the workplace, we explored the feasibility and accuracy of estimating spinal loads and kinematics during such tasks. A male participant was asked to lift and transfer a 10 kg load between two platforms. These platforms were positioned 190 cm apart, and each had a height of 30 cm from the floor. The task involved carrying the load from one platform to the other. After each transfer, the participant walked back to the starting platform and walked back to the other platform without the load, and the pattern was repeated six times in total. During the task, the subject carried the load in a free-style manner and performed the task at a self-selected pace. During the task whole-body kinematics were simultaneously recorded at 30 Hz through an eight-camera Optotrak system (Northern Digital Inc., Ontario, Canada). Further experimental details are described elsewhere (Muller et al., 2022). A single camera recording, parallel to the platform, also allowed the 3D pose estimation using MeTRAbs. Three metrics were considered to compare the kinematics during lifting: flexion angle, horizontal mid-hand to mid-shoulder distance (normalized to the shoulder width; representing load lever arm), and asymmetry angle. These parameters – obtained from both the Optotrak system and the MeTRAbs (Sárándi et al., 2023) pose estimation algorithm – were subsequently utilized in subject-specific static regression equations to estimate spinal loads (Ghezelbash et al., 2020).

### 2.2.3  Application III: Whiplash Neck Injury

Whiplash head and injury is particularly prevalent in forensic biomechanics applications. The injury risk assessment often relies on head/neck velocity and acceleration (Hayes et al., 2007); for example, some neck injury criteria involve head to neck relative acceleration and velocity (Hayes et al., 2007), and head impact power index (a metric for assessing severe head injuries) also requires head velocity and acceleration (Hayes et al., 2007). However, current pose estimation algorithms typically provide a static position only. Initially, we assessed the performance of MeTRAbs (Sárándi et al., 2023) in estimating the linear accelerations of the head (both vertical and horizontal) of a crash test dummy in a controlled lab environment (IIHS, 2023). For comparison with the pose estimation algorithm, we also estimated linear head accelerations (horizontal and vertical) by tracking the center of gravity symbol at head using template matching algorithm with a threshold of 0.8. Both pose estimation and template matching provided position data. Linear accelerations were computed using the finite difference technique with a second-order accuracy. Results were then filtered with a zero-lag low-pass Butterworth filter (order 5; frequency cutoff: 1.5 Hz). Additionally, in a real-world case of a whiplash injury during a rugby match, we utilized the available televised recordings of the game (Die Hard Rugby, 2023) and employed MeTRAbs (Sárándi et al., 2023) for pose estimation.

### *2.2.4 Application IV: Deadlift*

Video footage of a deadlift, 40 kg (Ethier, 2023), was used to estimate the 3D pose utilizing MeTRAbs (Sárándi et al., 2023). Since most pose estimation algorithms, including MeTRAbs, provide joint locations, we estimated the sacral rotation through lumbopelvic rhythm (Pries et al., 2015). These inputs – including trunk flexion angle, sacral rotation, and load location – were then considered to drive a subject-specific finite element musculoskeletal model of the spine (Ghezelbash et al., 2023) for a comprehensive biomechanical analysis. For more details on the musculoskeletal modeling, see Supplementary Materials and (Ghezelbash et al., 2023).

### *2.2.5 Application V: Snatch Weightlifting*

As an extreme lifting sports activity, we assessed spinal biomechanics during snatch weightlifting. The kinematics during snatching in two Olympic athletes – 222 kg (Weightlifting House, 2023) and 75 kg (Torokhtiy, 2023) – were estimated with MeTRAbs (Sárándi et al., 2023). Trunk flexion angle, arm angle, and hip displacement were computed and used to drive a dynamic subject-specific musculoskeletal model of the spine (Eskandari et al., 2023a, 2023b). Sacral rotation in these simulations was estimated using the lumbopelvic rhythm (Pries et al., 2015). For dynamic analyses, the velocity and acceleration components of the foregoing parameters were evaluated using a second-order finite difference approximation. Since the scaling algorithm to personalize the musculoskeletal model was originally developed for the average population, we increased the maximum muscle stress to 5 MPa to account for the likely differences with professional weightlifters (Ramirez et al., 2023).

## 3  Results and Discussion

In this study, we introduced a framework that synergistically integrates machine learning and computer vision technologies with musculoskeletal biomechanical methodologies to enable a comprehensive analysis of spinal biomechanics with essential input data from as simple as a single or multiple camera images. Unlike in a laboratory setting with complex measurement equipment and professional operators, the protocol only requires a single camera. This framework employs machine learning to estimate both 3D body shape and kinematics/pose from the single image (for static simulations) or multiple images (for dynamic simulations). This development paves the way for more affordable and practical applications in future biomechanical analyses, particularly in spine biomechanics. The following section presents and discusses the results of cases studied to evaluate the feasibility, accuracy, and potential of these approaches towards future opportunities and challenges.

## 3.1 Body Shape and Anthropometric Estimations

An accurate reconstruction of the body shape is essential for spine biomechanical analyses, especially in obese individuals where body shape can substantially alter segmental masses (e.g., arms and trunk) (Ghezelbash et al., 2017), Figure S1. Traditional methods to obtain 3D scans often require multiple view images or laser scans (Pryce and Kriellaars, 2014) or DEXA scans (Bachrach, 2000). In an attempt to simplify this process, we used PIFuHD (Saito et al., 2020) to reconstruct the 3D body shapes of four subjects in their T-poses from a single frontal image per subject. Rather small differences between estimated and actual measurements were found; the waist circumference and trunk mass differed by 10% (range: 1% to 15%) and 6% (range: -5 to 16%), respectively. While this approach seemed to work relatively well for lean male subjects, there were specific inaccuracies in the reconstructed 3D shapes, particularly with the discontinuous arms (Figure 2c) and fictitious holes (as marked by arrows in Figure 2a). Although this technology offers the potential to easily personalize musculoskeletal models without any medical images, and to collect body measurements without time-consuming traditional methods, further improvements and validations are necessary to overcome existing limitations.

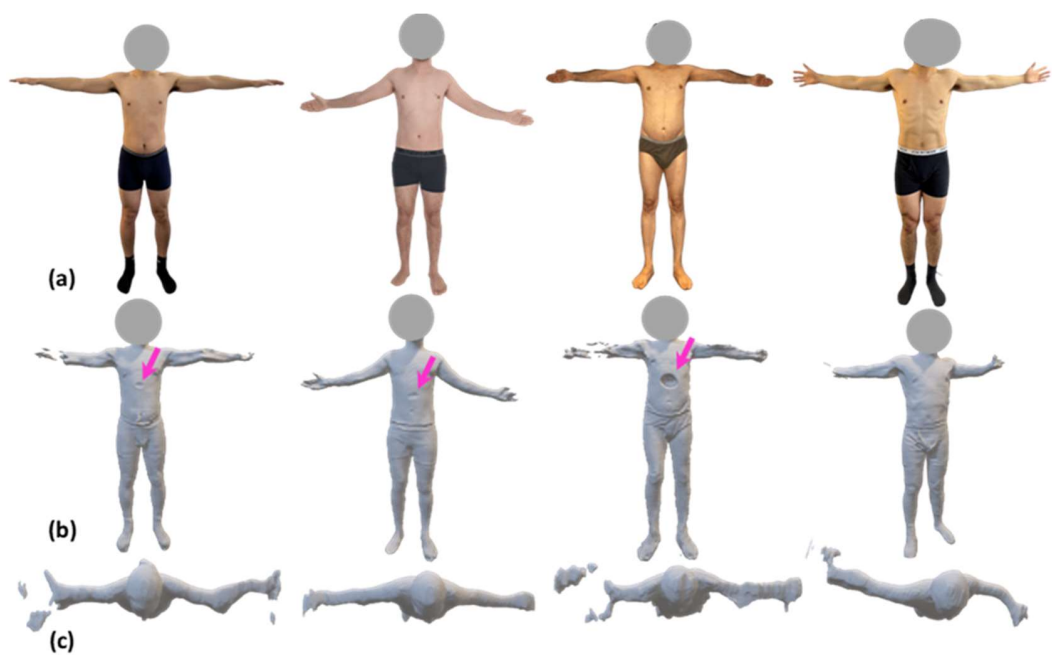

Figure 2: (a) Four lean subjects assuming a T-pose alongside their reconstructed 3D body scans shown in (b) frontal and (c) transverse planes. Defects in the reconstructed 3D shapes (b) are highlighted by magenta arrows.

## 3.2 Pose Estimation

As a cornerstone in most biomechanical analyses, kinematics have historically been collected by multi-camera/sensor systems. BlazePose was selected for its efficient and real-time processing capabilities. ICON was chosen for its unique ability to generate silhouette-type meshes, providing a more detailed representation of body contours, essential for in-depth biomechanical studies. We utilized MeTRAbs due to its status as the most updated and comprehensive algorithm, trained on a wide range of 23 datasets, ensuring both high robustness and accuracy in its estimations. We, however, explored the potential of a single-view image pose estimation using Blazepose (Bazarevsky et al., 2020), ICON (Xiu et al., 2022), and MeTRAbs (Sárándi et al., 2023) machine learning algorithms. Among the three, Blazepose's performance was found poor within the Fit3D dataset (Fieraru et al., 2021) (Figure S2) and failed in predicting 3D pose in real-world applications as it was conceived to be a lightweight pose estimator and not a tool with high accuracy. In contrast, ICON (Xiu et al., 2022), while delivering satisfactory results in Fit3D dataset (Fieraru et al., 2021), was time-consuming in pose estimation. MeTRAbs (Sárándi et al., 2023) stood out as the overall preferred package of choice in most applications due to its robustness, accuracy and reasonable computational demands. While there are many other algorithms such as DensePose, OpenPose, and AlphaPose, it is important to note that in comparison with MeTRAbs, they are often trained on more limited and older datasets.

### 3.2.1  Application I: Repetitive Lifting and Lyapunov Exponent

The application of Lyapunov exponents in clinical settings has been limited partly due to challenges in kinematics measurements. By analyzing a video recorded by a smart phone, ICON (Xiu et al., 2022) (Figure 3b) and MeTRAbs (Sárándi et al., 2023) (Figure 3d) successfully captured the trends in movements, yet they both underestimated the maximum whereas overestimated the minimum flexion angles in cycles (Figure 3). In some instances, ICON (Xiu et al., 2022) demonstrated irregular predictions (as indicated by arrows in Figure 3b). Using pose detection has yielded a trajectory reconstruction different from the measured one (Figure 3e). This is significant because time-delayed trajectory reconstruction (employed to reconstruct phase space when the full set of system variables is unknown) could potentially lead to a different understanding of system dynamics. Lyapunov exponents based on the estimated trunk flexion angle varied slightly from those derived from Xsens (0.51 versus 0.46; Figure 3f), indicating that the Lyapunov exponent depends on the divergence between adjacent trajectories and is not directly influenced by the absolute magnitude of the estimated flexion angle. However, it should be noted that pose estimation methods typically provide landmark points rather than rotations, thereby precluding our ability to estimate and incorporate the sacral rotation into the Lyapunov analysis or estimating 3D joint rotations from lateral images. Further studies on a larger population are warranted to assess the reliability of these techniques.

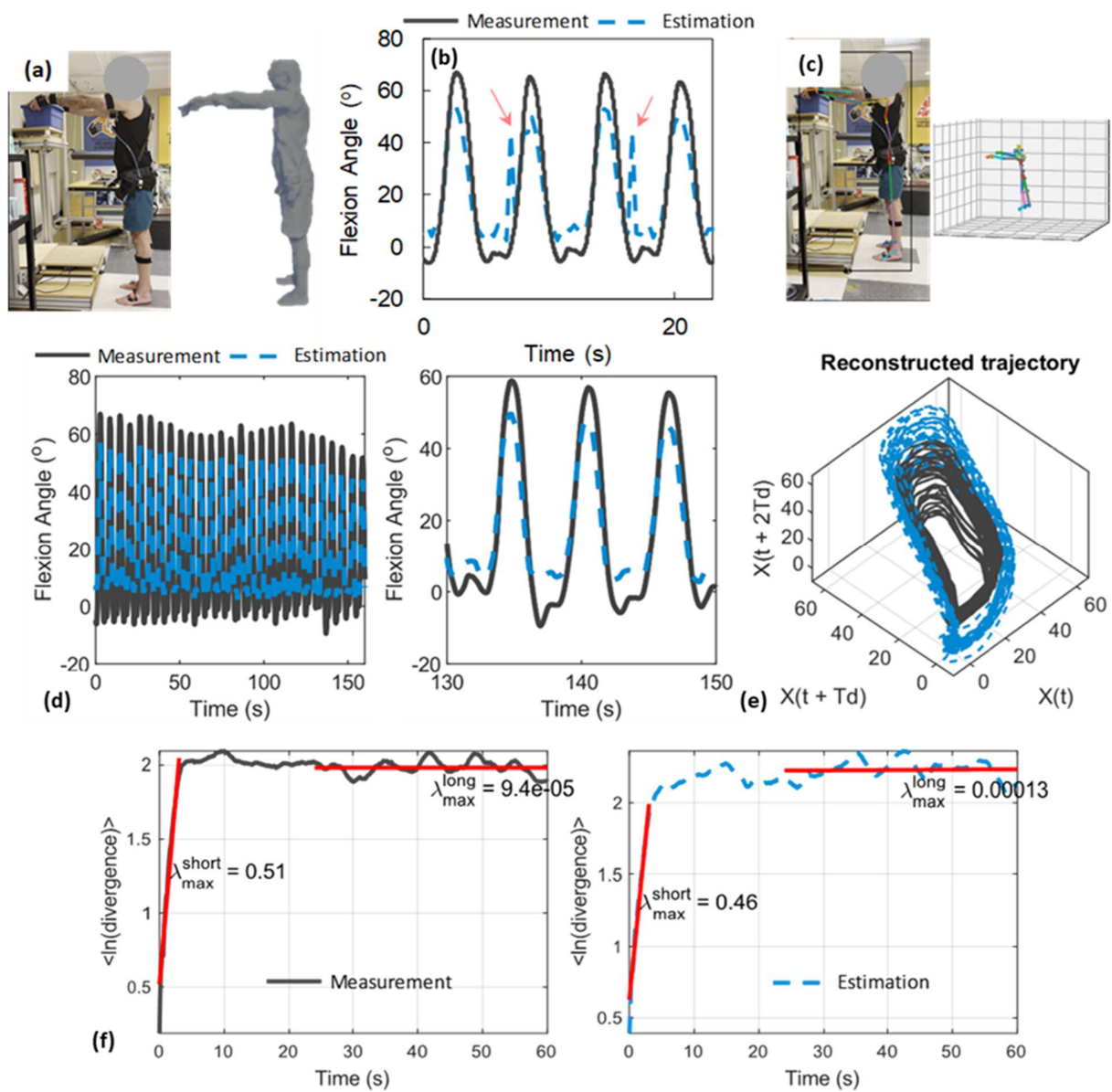

Figure 3: Parameters estimated during repetitive lifting, analyzed from smartphone videos and Xsens measurements: (a) Pose and (b) flexion angle estimated by ICON. (c) Pose and (d) flexion angle predictions using MeTRAbs. (e) Reconstructed trajectory (X, system state = flexion angle; Td, time delay = 0.6 s) for Xsens measurements and MeTRAbs estimations. (f) Calculated short and long Lyapunov exponents from Xsens measurements (left) and MeTRAbs estimations (right). Solid gray and dashed blue lines represent Xsens measurements and pose detection estimations, respectively.

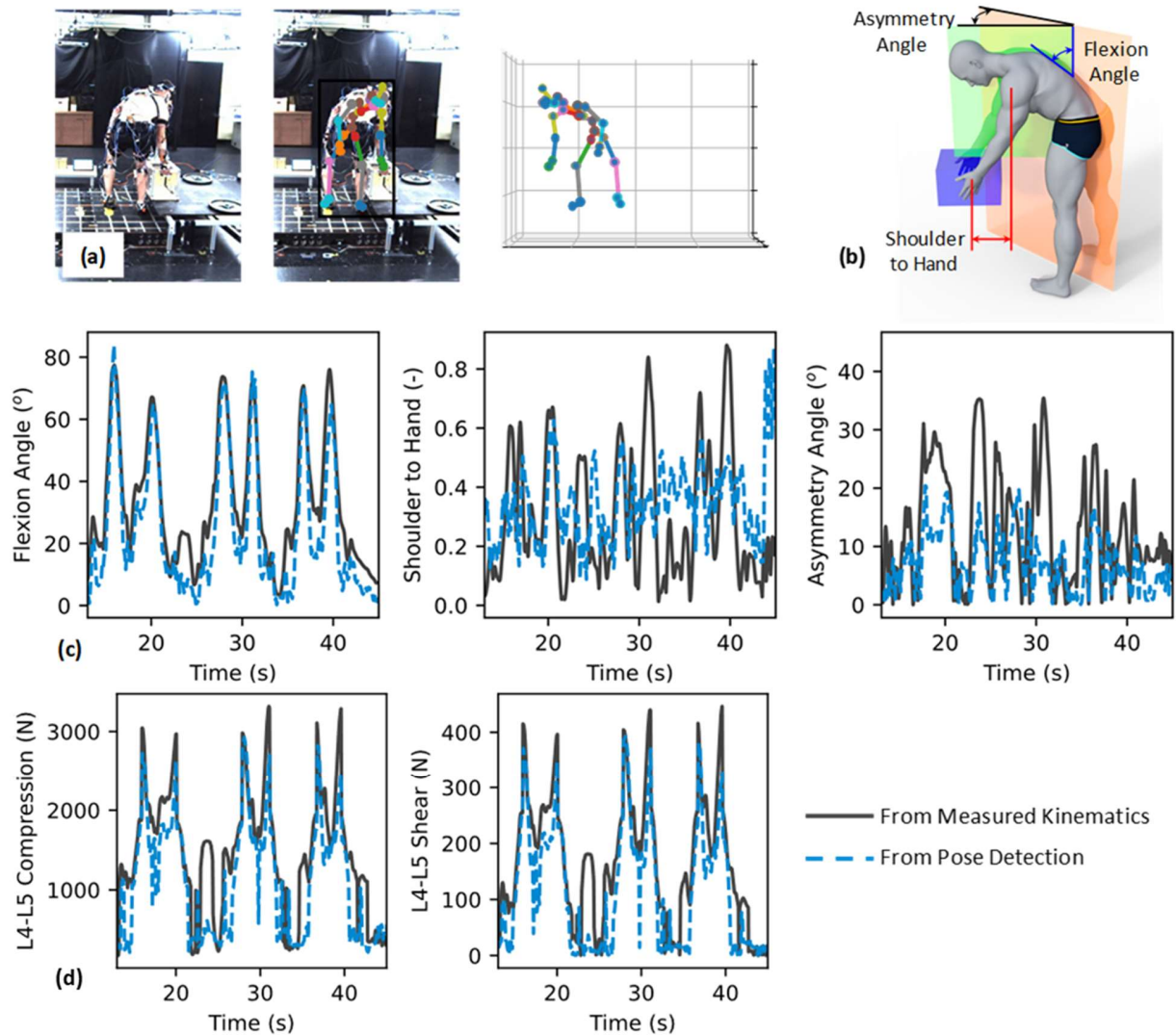

Figure 4: (a) An estimated pose during asymmetric lifting of a load. (b) Schematic representation of trunk flexion angle, shoulder-to-hand distance, as well as the asymmetry angle. (c) Estimated kinematics values from MeTRAbs versus measurements from the Optotrak system. Shoulder-to-hand horizontal distance was normalized to shoulder width, and therefore it is dimensionless. (d) Estimated L4-L5 shear and compression forces computed from subject-specific regression equations. Subject's height and weight were 176 cm and 72 kg.

### 3.2.2 *Application II: Asymmetric Lifting*

Various lifting assessment tools, such as NIOSH (Ghezelbash et al., 2020; Waters et al., 1994) and regression equations (Cholewicki et al., 1991), have been developed to help evaluate spinal loads and workplace safety during manual handling of loads; however, in-field applications of such tools have been limited because of kinematics measurements. Cameras, however, are widely available in workplaces to estimate kinematics with the goal to promote in-field biomechanical risk assessments and injury preventions. In comparison to Optotrak, MeTRAbs (Sárándi et al., 2023) estimated the flexion angle accurately, and captured the trend (but not the magnitude) of the asymmetry angle (Figure 4c). However, it failed to accurately predict the load horizontal distance from the shoulder (normalized to shoulder width); Figure 4c. Predicted spinal loads, nevertheless, agreed well with those based on Optotrak data (Figure 4d). It should, however, be noted that during some phases of the movement, the subject carried no load, whereas in others with a load held near the body, resulting in reduced moment arms. The asymmetry angle also remained relatively small during the entire movement. Such factors overall led to agreement between predicted spinal loads. Due to a rather poor prediction of the shoulder-to-hand horizontal distance (i.e., moment arm) and asymmetry angle, larger disagreement would be expected if greater loads were considered. Spinal loads were estimated using regression equations from a static musculoskeletal model (F Ghezelbash et al., 2020b). While dynamic analysis would ideally be preferable (Wang et al., 2021); since lifting tasks were performed at a slow to moderate pace, inertial and dynamic effects remain negligible (Bazrgari et al., 2008).

To utilize such technologies in industrial settings, some improvements are deemed necessary. First, the accuracy should be improved by exploring accurate/robust calibration methods and/or incorporating more training datasets. In addition, determining the load magnitude, as well as the interaction of the individual with the environment, are additional concerns. For instance, a wide range of activities (including pushing and pulling) could take place in the workplace. Furthermore, there is a need for more detailed evaluation of these interactions as well dynamic factors such as linear/angular velocities/accelerations along with mass and moment of inertia data.

### 3.2.3   Application III: Whiplash Neck Injury

The head acceleration is recognized as the crucial parameter when assessing the risk of whiplash injury (Hayes et al., 2007). In this context, MeTRAbs (Sárándi et al., 2023) demonstrated a performance comparable to the template matching method in predicting both vertical and horizontal linear accelerations of a crash test dummy (Figure 5a) although with noisier predictions. In another application, when MeTRAbs (Sárándi et al., 2023) was applied to a particular incident in a rugby match, it succeeded in identifying the distinct poses of both players during the early frames, when there was no contact between them (Figure 5c). Once contact initiated between the two players, MeTRAbs (Sárándi et al., 2023) failed to detect their individual poses (as indicated by arrows and asterisks in Figures 5 c,d). This illustrates a specific limitation of pose estimation algorithms, especially in dynamic and contact sport situations.

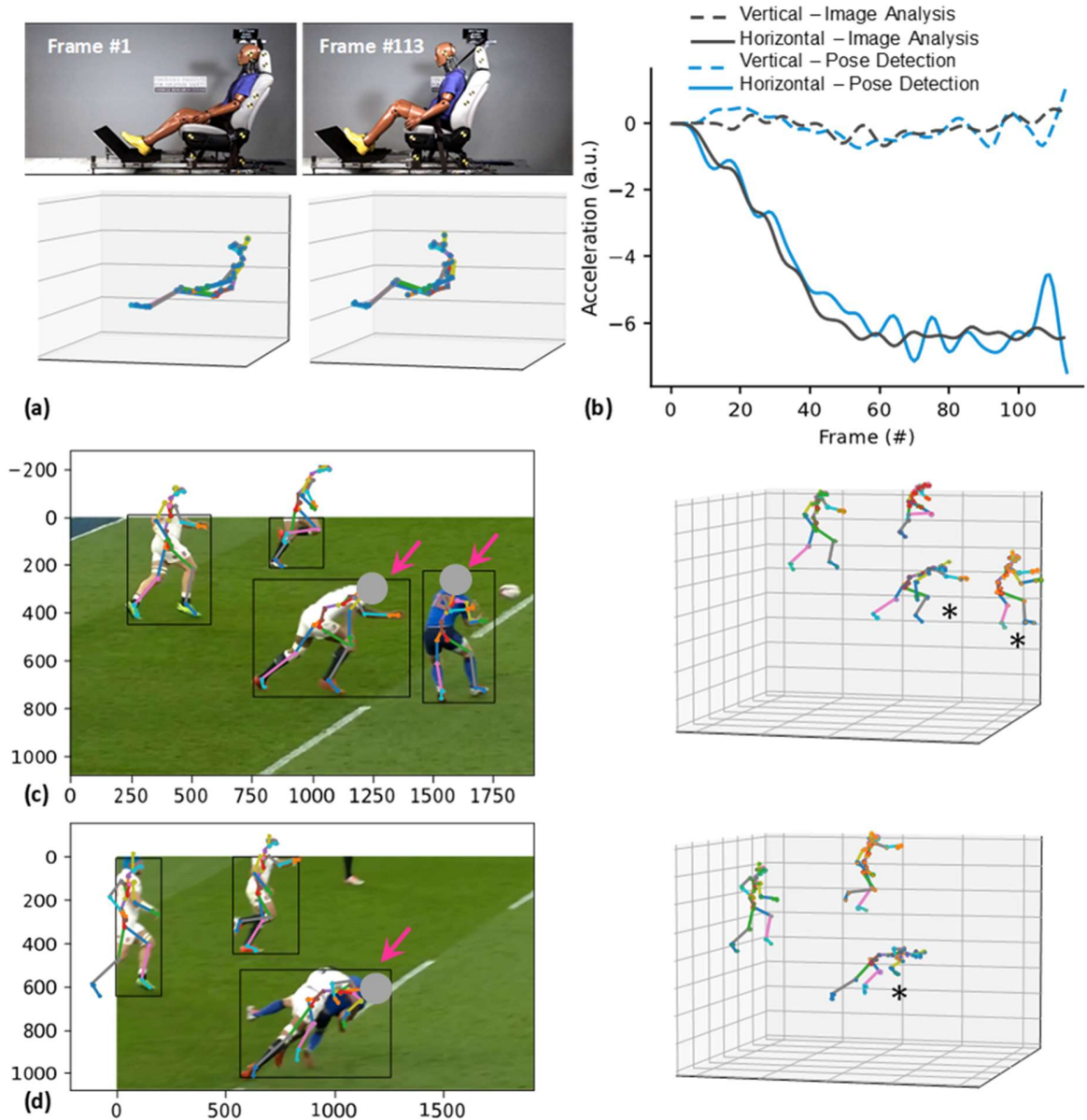

Figure 5: Estimated (a) poses as well as (b) head accelerations of a crash test dummy. When calculating the underline{acceleration}, we were unable to determine the frame rate and scale the length; therefore, the unit for acceleration is pixel/frame$^2$, an arbitrary unit (a.u.). Frames and estimated poses are shown (c) before and (d) during the contact in a rugby game. Individuals of interest are marked by arrows on the left and asterisks on the right.

### 3.2.4 Application IV: Deadlift

MeTRAbs (Sárándi et al., 2023) was employed to detect the 3D pose during a deadlift activity (Figure 6a). While athletes typically extend their spine during the final lifting stage of a deadlift, MeTRAbs predicted a 20° trunk flexion (Figure 6b). When incorporated in a coupled finite element-musculoskeletal model (Ghezelbash et al., 2023), large collagen fiber strains at the L4-L5 level, reaching up to 13% (at inner lateral/posterolateral layers; Figure 6c), were predicted. Despite the need for a more accurate prediction of kinematics, pose estimation algorithms such as MeTRAbs are found promising to pave the way for in-field biomechanical analyses – an accomplishment that would have seemed unfeasible only a few years prior.

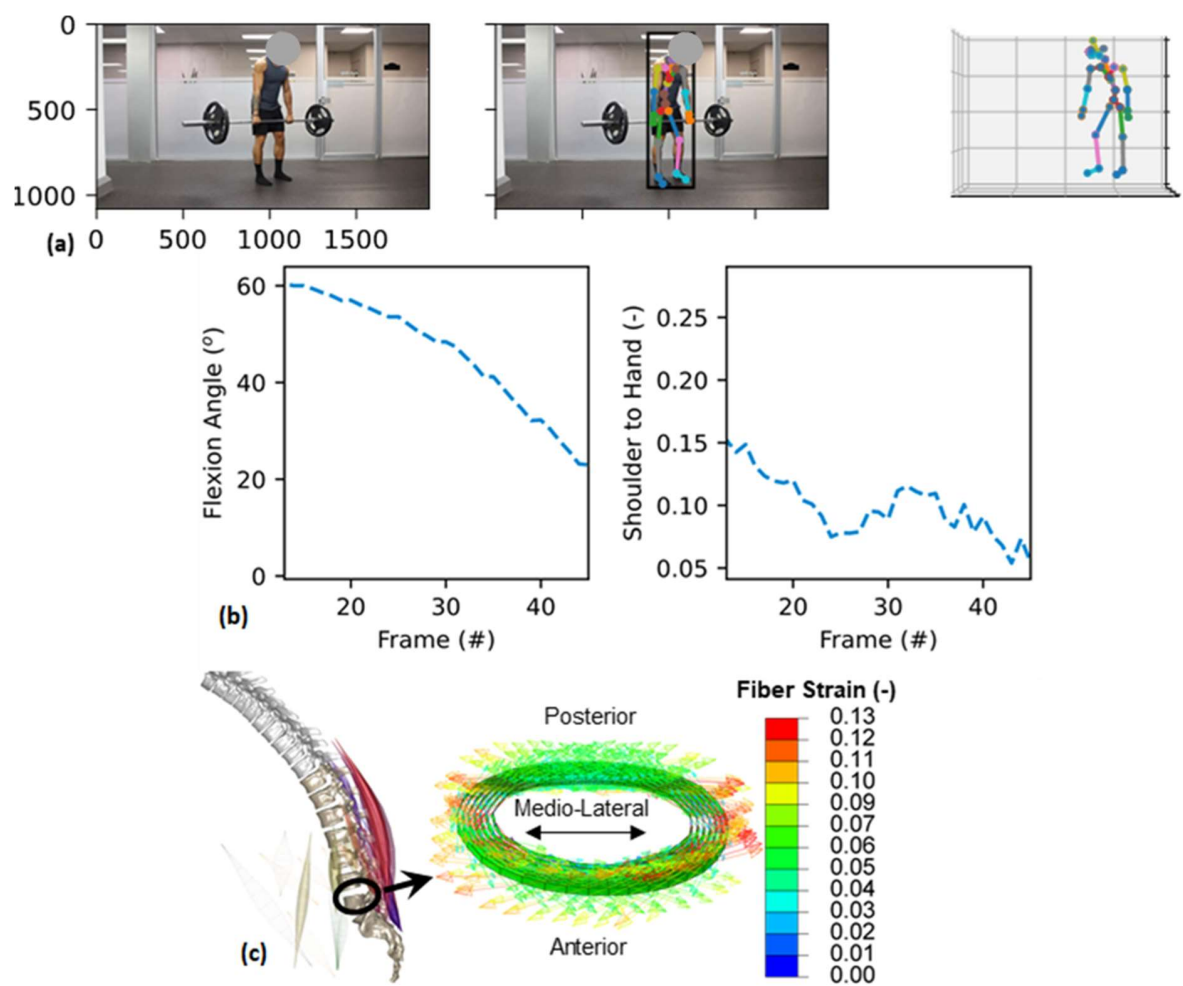

Figure 6: Estimated (a) pose, (b) trunk flexion angle, and shoulder-to-hand distance (normalized to the shoulder width) during deadlifting of 40 kg. (c) Predicted collagen fiber strains at 40° flexion angle, computed from our detailed finite element musculoskeletal model of the spine. In the musculoskeletal model, subject's height, weight and age were set at 175 cm, 72 kg and 24 years.

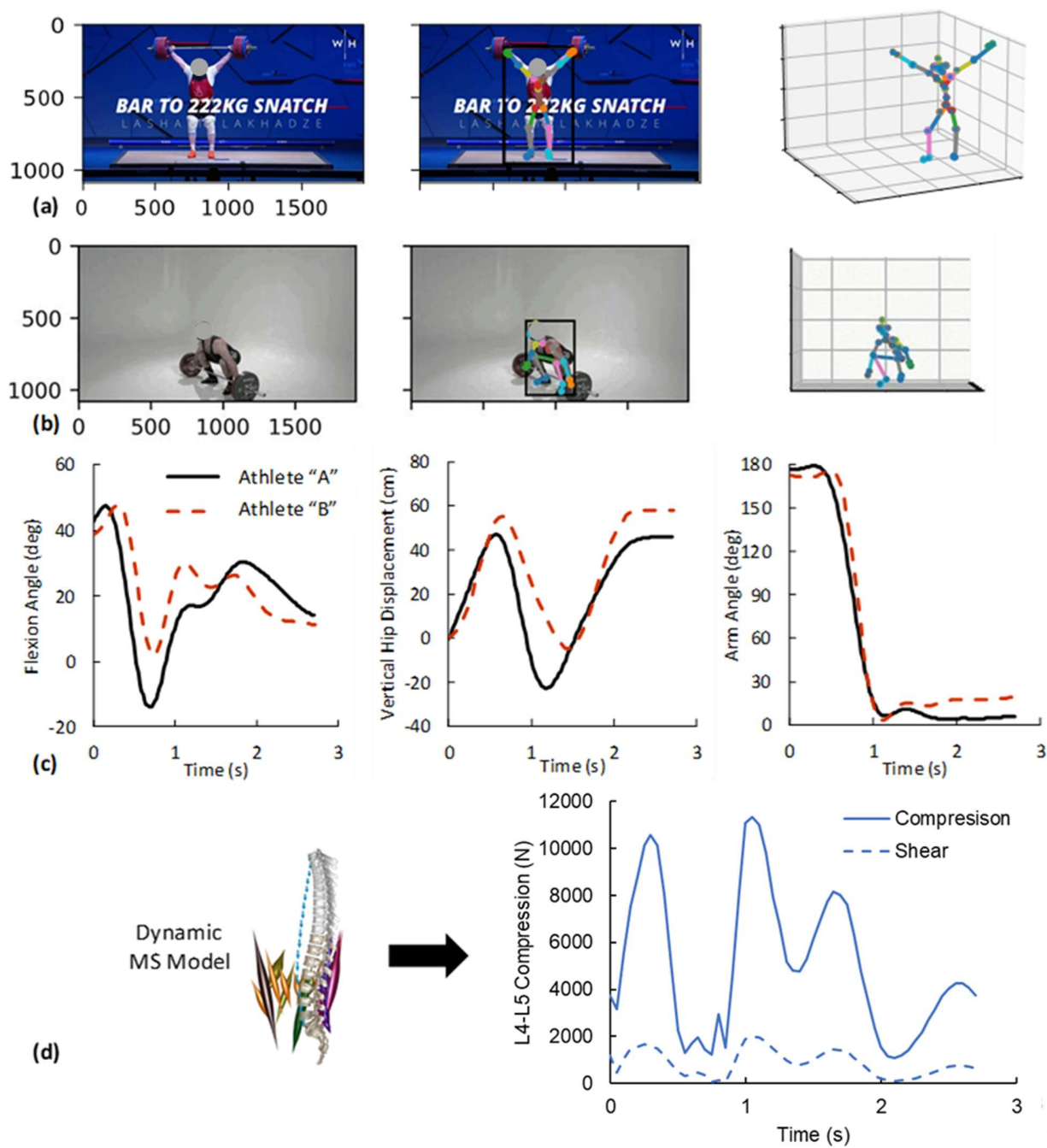

Figure 7: Snatching of (a) 222 kg by athlete "A" and (b) 75 kg by athlete "B" along with their estimated poses. (c) Estimated trunk flexion angle, vertical mid-hip displacement, and upper-arm angles during the snatch movement. (d) Estimated L4-L5 compression and shear, computed from a dynamic musculoskeletal model for Athleter "B".

### 3.2.5 Application V: Snatch Weightlifting

Within extreme sports, the study of snatch weightlifting technique is an interesting undertaking. In our study, we analyzed two professional weightlifters with a similar technique and kinematics employing MeTRAbs (Sárándi et al., 2023) (Figure 7), and both athletes exhibited a similar technique and kinematics during their performance (Figure 7c). Convergence issues prevented the simulation of the lifting of the 222 kg load. On the other hand, our analysis of the other athlete lifting a 75 kg load (by disregarding horizontal accelerations of the load) yielded large spinal compression at the L4-L5 level, reaching up to 11 kN, particularly during the initial upward lifting phase while holding the load (Figure 7d). These observations, achieved through machine learning algorithms and single-view footage, which underscore the potential of such technology, paving the way for in-field biomechanical analyses.

## 3.3  Potentials and Challenges

The integrated technology presents promising potential, especially in scenarios where traditional laboratory data collection may be impractical if not impossible. For instance, in forensic incidents where an accident has already occurred, employing these methods can facilitate biomechanical analyses to investigate the risk of injuries. In sports activities, the utilization of additional sensors might interfere with athletes' performance, so these technologies offer a less intrusive alternative. Occupational biomechanics, too, can benefit, as traditional approaches could impede normal function or cause additional effort for workers. The potential applications also extend into clinical biomechanics and remote health monitoring, broadening the scope and practical usage (Arrowsmith et al., 2022; Boswell et al., 2023).

While we demonstrated that these methods generally hold potential in various applications, it is also crucial to recognize that they do have several limitations demanding future research:

- Pose estimation approaches may violate physiological and biomechanical constraints, causing inaccuracies in the representation of the subject's anthropometry and movement.

- These methods could fail to distinguish individuals in contact, as illustrated in Application III where complex movements between individuals were not correctly captured. This is particularly crucial in sports and injury contexts where physical contact is common, highlighting a significant gap in current pose detection capabilities for accurate biomechanical analysis.

- Determining the interaction of the individual with the environment and the external loads is another concern. For example, in contrast to the lifting activities, automatically estimating the external load magnitude and direction (e.g., pull, push) and the exact instant of lifting in the field can present challenges.

- Although some studies have attempted to validate pose detection algorithms, our results have shown that there is a need to improve the overall accuracy, as some predictions of kinematic parameters did not align well with measurements. Optimal location of the camera, enhancements in calibration methods, and dataset diversity could be vital.

- Often, pose detection algorithms determine joint locations rather than the joint rotations. In this study, sacral rotation was estimated through lumbopelvic rhythm, pointing to a need for more advanced techniques to assess complex features such as joint 3D finite rotations.

- Single pose estimation focuses on stationary positions, missing out on the complexities of dynamic movements with velocities and accelerations. Adequately estimating mass and inertia can also be challenging for individuals and objects.

As this is a rapidly growing field, a more comprehensive exploration of newly developed methods, delving deeper into their performance, is crucial to fully realize their potential and address current constraints and limitations.

# 4 Acknowledgments

The study was partly supported by Natural Sciences and Engineering Council of Canada (NSERC RGPIN-03356).

# 5 Code Availability

The code used in our study, except for the musculoskeletal modeling, is publicly available at the following GitHub repository: https://github.com/farghea/Pose-Detection-Biomechanics. The musculoskeletal modeling code is available upon a reasonable request.

# 7 Supplementary Figures

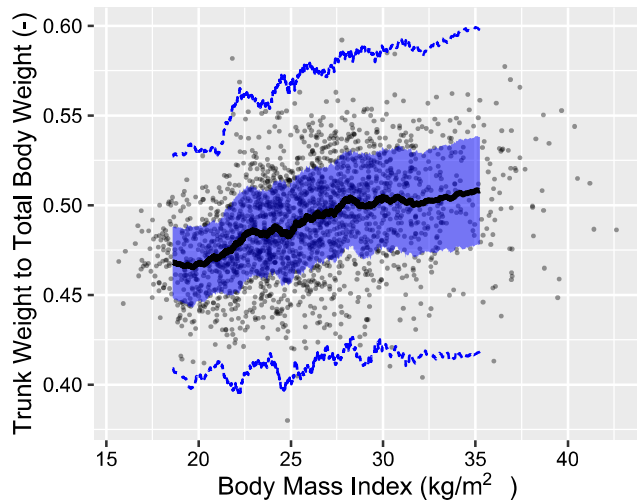

Figure S1: Variations in normalized trunk weight against body mass index

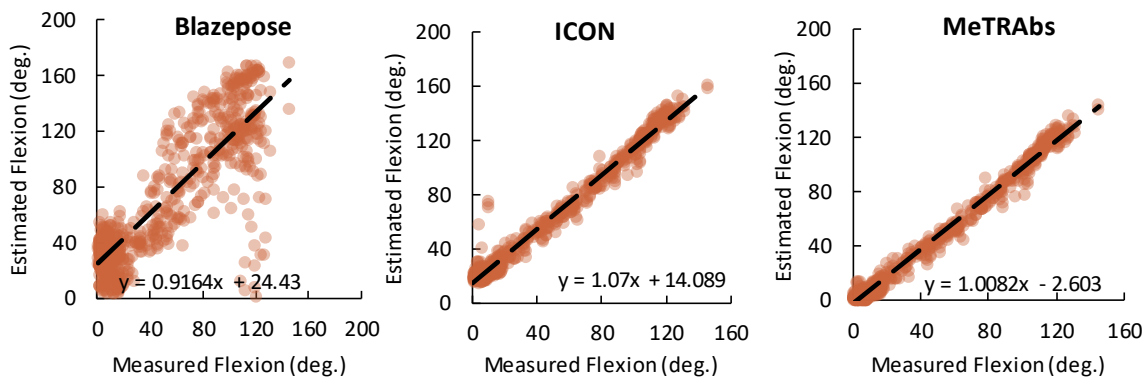

Figure S2: Measured versus estimated flexion angles for Blazepose, ICON and MeTRAbs in Fit3D dataset

Table S1: Measured versus estimated waist circumference (WC) for the subjects

| Subject # | Subject 1 | Subject 2 | Subject 3 | Subject 4 |
|---|---|---|---|---|
| Measured WC (cm) | 85 | 89 | 90 | 84 |
| Estimated WC (cm) | 86 | 102 | 100 | 96 |

Table S2: Correlation coefficient and absolute error (mean ± standard deviation) for measured versus estimated flexion angles using various pose detection algorithms

| Pose Detection Method | Correlation Coefficient | Absolute Error (°) |
|---|---|---|
| MeTRAbs | 0.96 | 7.2 ± 4.1 |
| ICON | 0.92 | 9.8 ± 7.3 |

Table S3: Correlation coefficient and absolute error (mean ± standard deviation) for measured versus estimated flexion angles, shoulder to hand distance, and asymmetry angle during asymmetric lifting

| Parameter | Flexion (°) | Shoulder to Hand (-) | Asymmetry Angle (°) |
|-----------|-------------|----------------------|---------------------|
| Correlation | 0.95 | 0.16 | 0.50 |
| Absolute Error | 8.8 ± 5.7 | 0.19 ± 0.11 | **7.8 ± 6.5** |

## 7.9  Finite Element Subject-Specific Musculoskeletal Model

A detailed finite element (FE) model of the lumbar spine (T12 to S1) was used as the reference model (Ghezelbash et al., 2023), and a scaling algorithm was introduced to individualize this FE model of the spine using regression equations and statistical models (Ghezelbash et al., 2023). To individualize the model, we parametrized the reference model (Figure S3) based on existing imaging studies and statistical shape models in accordance with subject's sex, age, height and weight (Tang et al., 2022, 2019). Disc areas (anterior-posterior and medio-lateral disc diameters) were adjusted based on a subject's sex and height (Tang et al., 2019); Figure S3, and segmental lordosis was scaled in accordance with a statistical shape model of the lumbar spine (Tang et al., 2022), Figure S3. Disc heights were assumed to be proportional to the subject's body height. Nodal coordinates of posterior elements (e.g., facet articular surfaces, transverse processes, spinous processes) were modified proportional to the disc geometry (Figure S3). Collagen fiber network parameters (i.e., fiber area and angle) were adjusted to maintain the same volume fraction content and orientation within the respective intervertebral disc.

The foregoing individualized FE model of the passive spine was integrated with a comprehensive musculature incorporating 126 sagittally-symmetric muscles (Eskandari et al., 2023; Ghezelbash et al., 2023, 2016). Muscle moment arms and cross-sectional areas were individualized with subject's anthropometric parameters (i.e., sex, age, height and weight) (Anderson et al., 2012); Figure S3. Segmental masses along the trunk height were proportionally scaled based on the subject's body weight while their locations were adjusted based on the subject's body height (De Leva, 1996; Pearsall et al., 1996). In each task and after the individualization of the passive spine model, musculature and segmental masses; measured kinematics (i.e., segmental rotations at T12 to S1) and external loads were applied into the model, and muscle forces were estimated (using the minimization of the sum of squared muscle stresses constrained by equilibrium equations at all levels-directions). Muscle forces were updated and added to the existing external loads in the next iteration. The analysis was repeated until convergence was reached (<5% changes in muscle forces) (Ghezelbash et al., 2023; Seth and Pandy, 2007).

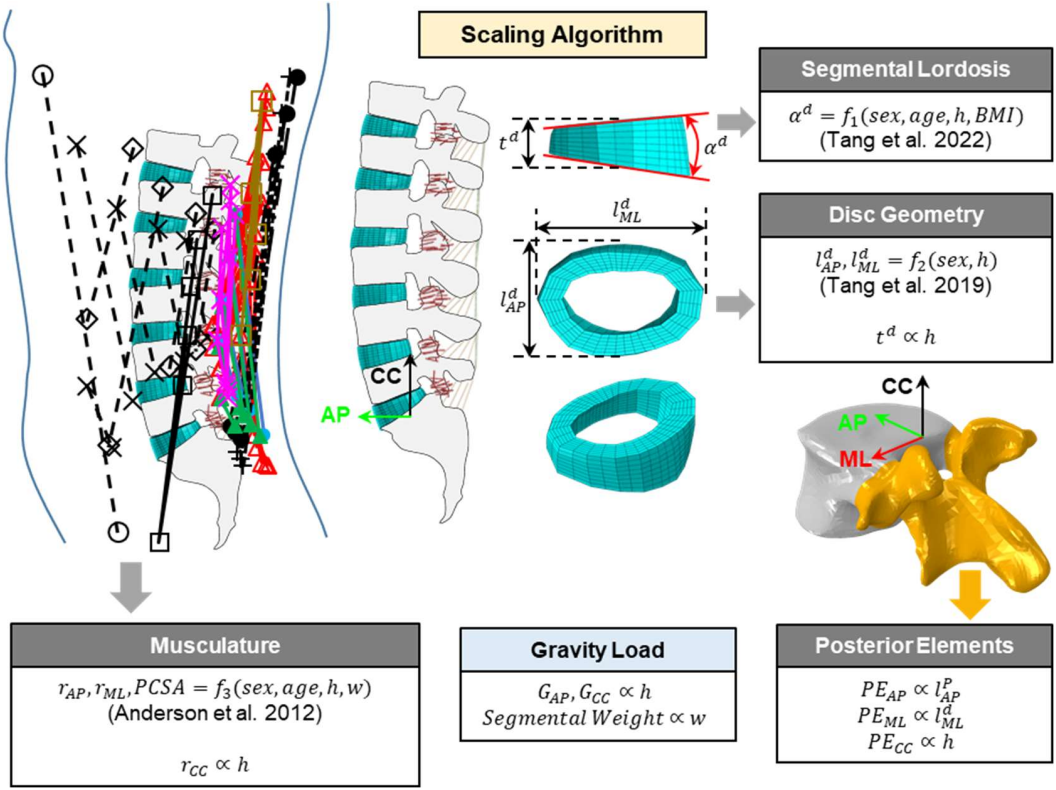

Figure S3: Schematics of the scaling algorithm and the subject-specific finite element musculoskeletal model ($\alpha^d$: segmental lordosis; $t^d$: mean disc height; $l^d$: disc diameter; $PE$: nodal coordinates of posterior elements; $M$: muscle moment arm; $PCSA$: physiological cross sectional area; $G$: gravity load location; $h$: body height; $w$: body weight; $BMI$: body mass index; $AP$: anterior-posterior; $ML$: medio-lateral; $CC$: cranial-caudal).